# Absence of zero-point entropy in a triangular Ising antiferromagnet

Yuesheng Li[1], Sebastian Bachus[1], Yoshifumi Tokiwa[1], Alexander A. Tsirlin[1] & Philipp Gegenwart[1]

[1]Experimental Physics VI, Center for Electronic Correlations and Magnetism, University of Augsburg, 86159 Augsburg, Germany.

**Frustrated Ising magnets host exotic excitations, such as magnetic monopoles in spin ice[1-4]. The ground state (GS) in this case is characterized by an extensive degeneracy and associated residual entropy going back to the pioneering work by G. Wannier who established large residual entropy of nearly 50%$R$ln2 per mole spins in a triangular Ising antiferromagnet (TIAF) already in 1950[5-7]. Here, we endeavor to verify this result experimentally using $TmMgGaO_4$, a novel rare-earth-based frustrated antiferromagnet with Ising spins arranged on a perfect triangular lattice. Contrary to theoretical expectations, we find almost no residual entropy and ascribe this result to the presence of a weak second-neighbor coupling $J_2^{zz} \sim 0.09 J_1^{zz}$ that lifts the GS degeneracy and gives rise to several ordered states, the stripe order, 1/3-plateau, and 1/2-plateau. $TmMgGaO_4$ gives experimental access to these novel phases of Ising spins on the triangular lattice.**

Geometrical frustration renders the GS macroscopically degenerate in some spin systems and gives rise to a significant zero-point entropy in contradiction to the third law of thermodynamics[1]. One of the most extensively studied systems of this type is the pyrochlore Ising ferromagnet, $Dy_2Ti_2O_7$, which mimics the disordered proton arrangement in water ice and Pauling's ice rule[2,3]. The spin-ice GS is macroscopically degenerate with a finite zero-point entropy, $S_0^p = (R/2)\ln(3/2) \sim 29\%R\ln2$[2,3], and with the exotic excitations of Dirac strings and magnetic monopoles[4]. Another important geometrically frustrated spin system is the TIAF, where the GS had been reported to show an even larger zero-point entropy, $S_0^t = 44\% - 50\% R\ln2$[5-7]. Whereas the spin-ice physics is nowadays well exemplified by rare-earth pyrochlores, no real-world prototype of the TIAF model has been reported to date.

In a search for such a material, we explored structural siblings of YbMgGaO$_4$, which we recently established as a quantum spin liquid candidate on the undistorted triangular lattice[8-12]. In this compound, magnetic moments of Yb$^{3+}$ are somewhat anisotropic, but the in-plane (*ab*-) component $g_\perp = 3.060(4)$ is still comparable to the out-of-plane (*c*-) component $g_\parallel = 3.721(6)$[9,13-15]. This clear deviation from the Ising regime may be linked to the Kramers nature of the Yb$^{3+}$ ion with the symmetry-protected GS doublet. A stronger Ising nature is expected in non-Kramers ions if their two low-lying singlet states (a quasidoublet) are well separated from other crystal electric field (CEF) levels[16]. This renders TmMgGaO$_4$ with the non-Kramers Tm$^{3+}$ a promising candidate for studying the TIAF physics. A very recent work by Cevallos *et al.*[17] demonstrated strong Ising nature of Tm$^{3+}$ spins in this compound indeed, but only basic measurements down to 1.8 K were reported, and residual spin entropy has not been probed.

Below, we report a thorough single-crystal investigation on the GS magnetism of the TIAF compound, TmMgGaO$_4$, including heat capacity, Faraday force magnetization & susceptibility, and magnetocaloric effect measurements down to 30 mK. TmMgGaO$_4$ shows a robust Ising anisotropy along the *c*-axis and a relatively strong nearest-neighbor (NN) antiferromagnetic coupling, $J_1^{zz} \sim 10$ K. However, almost no zero-point entropy, $S_0 \leq 0.6\% R\ln2 \ll S_0^t$, is observed at $\sim 0.1$ K, inconsistent with the pure NN TIAF model. By exploring magnetic transitions and phase diagram in the longitudinal field, we found that a non-negligible next-nearest-neighbor (NNN) interaction, mainly induced by the magnetic dipole-dipole interaction, $J_2^{zz} \sim 0.09 J_1^{zz}$, is large enough to release all zero-point entropy and stabilize a frozen, presumably stripe state below $\sim 0.27$ K. The applied magnetic field along the *c*-axis induces novel 1/3-plateau, 1/2-plateau, and spin-polarized phases consecutively at low temperatures, which is consistent with the random $J_1^{zz}$ - $J_2^{zz}$ TIAF model.

**Results**

**Effective pseudospin-1/2 Hamiltonian.** TmMgGaO$_4$ has a crystal structure ($R\bar{3}m$ space group) very similar to that of YbMgGaO$_4$[8,9,17], with the rare-earth Tm$^{3+}$ ions forming a two-dimensional triangular lattice. In TmMgGaO$_4$, the local CEF at the Tm$^{3+}$ site with the $D_{3d}$ point-group symmetry splits the thirteenfold-degenerate GS of the free Tm$^{3+}$ ion with the total angular momentum $J = 6$ and Landé $g$-factor $g_J = 7/6$, $|m_J\rangle$ ($m_J = 0, \pm 1..., \pm J$), into five singlets ($3A_{1g}+2A_{2g}$) and four doublets ($4E_g$), according to the symmetry analysis.

At low temperatures, the magnetization of TmMgGaO$_4$ shows a robust Ising anisotropy along the $c$-axis (see Fig. 1a). The magnetization perpendicular to the $c$-axis, $M_\perp$, is one order of magnitude smaller than $M_\parallel$, and is almost linearly field-dependent up to 7 T at 1.9 K, consistent with the concurrent report by Cevallos *et al*[17]. This suggests the presence of Ising spins. To better understand their nature, we prepared highly diluted samples of Tm$_x$Lu$_{1-x}$MgGaO$_4$ ($x = 0.04$), similar to our recent study of the spin-chain system of PrTiNbO$_6$[18]. The highly-diluted Yb$_x$Lu$_{1-x}$MgGaO$_4$ ($x = 0.04$) with Yb$^{3+}$ as Kramers ion was also studied as reference. In both bases, the dilution eliminates any intersite magnetic couplings, as revealed by the diminutively small Curie-Weiss temperatures, $\theta_w(x = 0.04) \sim 0.16\theta_w(x = 1)$ (see Fig. 1c for Tm$_x$Lu$_{1-x}$MgGaO$_4$ and Ref. 8 for Yb$_x$Lu$_{1-x}$MgGaO$_4$). The difference between the Kramers and non-Kramers cases is clearly seen in $C_m/T$, where the signal of the diluted Yb$^{3+}$ sample diverges at low temperatures, while the diluted Tm$^{3+}$ sample reveals a finite zero-temperature limit of $C_m/T$. This finite value indicates a distribution of the two low-lying CEF levels, $|E_1\rangle$ and $|E_2\rangle$, and their accidental degeneracy. Following the framework developed for PrTiNbO$_6$, we model this distribution with a Lorentzian function centered at $\Delta_0^{2,1} = \langle E_2-E_1\rangle$ and having the full width at half maximum (FWHM) $\omega$. The non-zero $\omega$ arises from the site mixing of Mg$^{2+}$ and Ga$^{3+}$ that, with their different charges, generate random CEF on the rare-earth site, an effect integral to the putative spin-liquid physics of

YbMgGaO$_4$[15]. By fitting $C_m/T$ of the diluted samples (Fig. 1d), we find $\Delta_0^{2,1}$ = 5.9 K and $\omega$ = 5.3 K for Tm$^{3+}$ compared to $\Delta_0^{2,1}$ = 0 and $\omega$ = 0.19 K for Yb$^{3+}$, where the GS doublet is protected by time-reversal symmetry. Whereas this protection does not occur in the case of Tm$^{3+}$, a robust quasidoublet forms, because $\omega$ is nearly as large as $\Delta_0^{2,1}$. This gives rise to the low-temperature magnetism even at temperatures well below $\Delta_0^{2,1}$. It is also worth noting that the zero-temperature limit of $C_m/T$ is nearly zero for pure TmMgGaO$_4$ at odds with the highly diluted sample, where the value is finite. Therefore, interactions between Ising spins open a gap in the spectrum of TmMgGaO$_4$.

At 1.9 K and above 10 T, the magnetization of TmMgGaO$_4$ shows a full polarization along the $c$-axis (see Fig. 1a), with a small Van Vleck susceptibility $\chi_\parallel^{vv}$ = 0.003(1) cm$^3$/mol Tm, and with an average pseudospin-1/2 $g$-factor $g_\parallel$ = 13.18(1), which is close to the upper limit of $2Jg_J$ = 14 for Tm$^{3+}$, indicating the nearly classical CEF GS quasidoublet mainly formed by the $|\pm6\rangle$ states. Assuming pure classical Ising nature of the pesudospins, an effective NN Hamiltonian for TmMgGaO$_4$ reads as[19,20],

$$\mathcal{H} = J_1^{zz} \sum_{\langle ij \rangle} S_i^z S_j^z. \qquad (1)$$

Through the Curie-Weiss fit to the susceptibility along the $c$-axis between 30 and 60 K (see Fig. 1c), we obtain an effective moment of $\mu_{\text{eff}}$ = 6.5(1)$\mu_B$ ~ $g_\parallel\mu_B/2$ and $\theta_w$ = -16.44(3) K. And we further get $J_1^{zz}$ ~ -2$\theta_w$/3 ~ 10 K.

**Absence of zero-point entropy.** The magnetic heat capacity ($C_m$) of TmMgGaO$_4$ can be determined accurately by subtracting the heat capacity of the non-magnetic LuMgGaO$_4$ as phonon contribution (Supplementary Fig. 1)[8,9]. We further obtain the magnetic entropy by integrating $C_m/T$ over $T$. The magnetic entropy of TmMgGaO$_4$ shows a broad plateau of $R$ln2 between 30 and 60 K, and sharply decreases down to ~ 0.6%$R$ln2 at 0.1 K (see Fig. 1b), confirming the effective pseudospin-1/2 physics below ~ 60 K[9,18].

Surprisingly, at ~ 0.1 K and 0 T the residual electronic spin entropy of TmMgGaO$_4$ is measured to be almost zero (see Fig. 1b), $S_m \leq$ 0.6%$R$ln2 $\ll S_0^t$, which conforms to the third law of thermodynamics. Whereas the heat capacity is smooth down to the lowest temperature of our measurement, a divergence of the field-cooled and zero-field-cooled susceptibilities indicates spin freezing below $T_c \sim 0.27$ K at 0.1 T (see Fig. 1e). These low-$T$ observations clearly conflict to the NN TIAF model of Eq. (1), which predicts the macroscopically degenerate GS[5-7]. Therefore, other interactions or perturbations must be taken into account to fully understand the low-$T$ magnetism of TmMgGaO$_4$. We explore them by studying thermodynamic properties in longitudinal magnetic field.

**Low-$T$ thermodynamic properties.** The magnetization measured at low temperatures shows interesting features (see Fig. 3a). After taking derivative with respect to the field, magnetic susceptibility is obtained (see Fig. 3b). At 2 K, the susceptibility shows two very broad humps at $\mu_0 H_h \sim 0.4$ and $\sim 3.6$ T respectively, which is consistent with the 1.8 K measurement of Ref. 17. At 0.2 K, these peaks become much sharper, and a new one appears at $\mu_0 H_c \sim 2.6$ T (see Fig. 3b). Further cooling to 40 mK has little effect, even though temperature and thermal fluctuations decrease by a factor of 5. Around the transition fields, the corresponding peaks are also clearly observed in the magnetic Grüneisen ratio (see Fig. 3c) and heat capacity (see Fig. 3d) measurements. At 0.3 K and above $\sim 3.6$ T, the spin system of TmMgGaO$_4$ is almost fully polarized, so it contributes little to the heat capacity (see Fig. 3d).

Three field-induced transitions are unexpected in the NN TIAF model, where only two field-induced states, the 1/3-plateau and fully polarized, should occur[21]. On the other hand, adding the NNN coupling $J_2^{zz}$ allows for additional field-induced phases and may explain the occurrence of three transitions[22]. In the following, we use the modified Hamiltonian,

$$\mathcal{H} = J_1^{zz} \sum_{\langle ij \rangle} S_i^z S_j^z + J_2^{zz} \sum_{\langle\langle kl \rangle\rangle} S_k^z S_l^z - \mu_0 H_{||} g_{||} \mu_B \sum_i S_i^z. \qquad (2)$$

to model the magnetization process of TmMgGaO$_4$.

Phenomenologically, the susceptibility at 40 mK can be well fitted by three Lorentzian peaks (see Fig. 3b). This way, three transition fields and the associated changes in the magnetization are determined. On the other hand, the broadened nature of the transitions is not captured by Eq. (2), as the calculated magnetization curve at 0 K should be step-like ($M_\parallel/M_\parallel^s$ = 0, 1/3, 1/2, or 1, see Fig. 2 and 3a), and three delta-peaks should be seen in the derivative. The broadening can not be caused by thermal fluctuations, as $T$ = 40 mK is two orders of magnitude smaller than the energy scale of the broadening (> 0.8 T). Moreover, no significant differences are observed between the 0.3 K and 40 mK data (see Fig. 3b). Therefore, a distribution of $g_\parallel$ and magnetic couplings $J_1^{zz}$ & $J_2^{zz}$ due to structural randomness should be taken into account, and a much better agreement is indeed achieved by assuming Lorentzian distributions of these three parameters (see Fig. 3a). We thus obtain $g_\parallel$ = 12.6 (FWHM = 1.5), $J_1^{zz}$ = 9.3 K (FWHM = 2.4 K), and $J_2^{zz}$ = 0.88 K (FWHM = 0.9 K) (Supplementary Fig. 2). The distribution of the parameters explains the smooth nature of the transitions even at the temperature of 40 mK. The interaction $J_2^{zz}$ is close to the magnetic dipole-dipole interaction limit, $\mu_0 g_\parallel^2 \mu_B^2/(4\pi r_{NNN}^3)$ ~ 0.52 K ($r_{NNN}$ = $\sqrt{3}a$ = 5.92 Å), suggesting only a small exchange contribution of ~ 0.3 K to the NNN coupling[9,17,18,23,24]. In the case of $J_1^{zz}$, the dipole-dipole interaction of $\mu_0 g_\parallel^2 \mu_B^2/(4\pi a^3)$ = 2.7 K accounts for less than one third of the total coupling. The quick decrease of the exchange contribution from ~ 6.6 K for nearest neighbors to ~ 0.3 K for next-nearest neighbors confirms the strongly localized nature of the 4$f$ electrons in TmMgGaO$_4$.

**Phase diagram.** To obtain the detailed low-$T$ phase diagram for TmMgGaO$_4$, we further measured temperature dependence of the heat capacity at different magnetic fields (see Fig. 4a). Below ~ 0.3 K, the magnetic heat capacity shows a robust Schottky tail with a total entropy of ~ 5.6%$R$ln2, which should originate from the $^{169}$Tm nuclear spins hyperfine-coupled to the local electronic magnetization. Between 0 to 0.2 T, the fitted energy gap for the Schottky tail increases from 56.7(2) to 63.0(1) mK (see Fig. 4a), with an effective gyromagnetic ratio, $\gamma_{eff}/(2\pi) = k_B[^{169}\Delta(H_\parallel) - ^{169}\Delta_0]/(\mu_0 H_\parallel h)$ = 660(30) MHz/T, which is two orders of magnitude smaller than the gyromagnetic ratio of free electrons and excludes the possible electronic spin origin, e.g., from free (defect) Tm$^{3+}$ electronic spins in general[18]. The local susceptibility, $\chi_\parallel^{loc} \sim N_A \mu_0 \mu_B g_J \gamma_{eff}/(2\pi |A_J|)$

~ 14(1) cm$^3$/mol Tm[18], is very close to the low-$T$ bulk susceptibility (see Fig. 1e), where |$A_J$| = 394 MHz is the hyperfine coupling of $^{169}$Tm$^{3+}$, two orders of magnitude larger than the gyromagnetic ratio of free $^{169}$Tm, $^{169}\gamma/(2\pi)$ = 3.5 MHz/T[25].

In zero field, two broad humps are observed in the magnetic heat capacity at $T_h$ ~ 0.9 K and ~ 2 K (see Fig. 4a). Under magnetic field up to ~ 1.5 T, a very sharp λ-peak at $T_c$ ~ 1 – 2 K appears. Upon further increase in the field, the sharp peak gradually becomes a broad hump again (see Fig. 4a and Supplementary Fig. 3). The transition is most sharp at ~ 1.5 T ($T_c$ ~ 1.6 K), it should be mainly driven by the strongest NN coupling $J_1^{zz}$. The low-$T$ magnetic phase diagram of TmMgGaO$_4$ is summarized in Fig. 4b. According to the earlier study of the $J_1^{zz}$ - $J_2^{zz}$ TIAF model on the triangular lattice[22], we identify the zero-field state as stripe order (affected by spin freezing), whereas the field-induced phases are the 1/3-plateau, 1/2-plateau, and the fully polarized state (see Fig. 2).

**Discussion**

We have shown that the random $J_1^{zz}$ - $J_2^{zz}$ TIAF model captures main features of the low-$T$ magnetism of TmMgGaO$_4$:

1) The increase in the magnetization around the transition fields equals to 0.37, 0.17, and 0.46$M_\parallel^s$ according to the areas of Lorentzian peaks in Fig. 3b between 0 and 6 T. These values are consistent, respectively, with 1/3, 1/6, and 1/2$M_\parallel^s$ expected for the classical $J_1^{zz}$ - $J_2^{zz}$ TIAF model (see Fig. 2b-e)[22]. Here, $M_\parallel^s$ = $g_\parallel\mu_B$/2 is the saturated magnetization.

2) The transition fields deliver consistent estimates of model parameters for the classical $J_1^{zz}$ - $J_2^{zz}$ TIAF Hamiltonian, Eq. (2): $\mu_0H_{c3}$ is compatible with the measured Curie-Weiss temperature (Fig. 1c), 3($J_1^{zz}$+$J_2^{zz}$)/($\mu_Bg_\parallel$) ~ -2$k_B\theta_w$/($\mu_Bg_\parallel$) in the mean-field approximation (see Fig. 3). The NNN coupling $J_2^{zz}$ ~ 0.7 K determined from $\mu_0(H_{c3}-H_{c2})$ = 12$J_2^{zz}$/($\mu_Bg_\parallel$) is close to the magnetic dipole-dipole interaction limit, $\mu_0g_\parallel^2\mu_B^2/(4\pi r_{NNN}^3)$.

3) At ~ 0 T, the transition temperature determined from susceptibility measurements, $T_c$ ~ 0.27 K (see Fig. 1e), and the position of the lower temperature hump in the heat capacity, $T_h \sim 0.9$ K (see Fig. 4a), are comparable to the median value of $J_2^{zz}$, which supports our hypothesis that spin freezing toward the stripe state (see Fig. 2b) is driven by $J_2^{zz}$.

4) By fitting the low-$T$ part of zero-field $C_m$-$C_n$ with a power-law function, $C_m \sim T^\gamma$, we arrive at a large exponent of $\gamma = 2.60(1)$ that exceeds $\gamma = 2$ in an ordered two-dimensional antiferromagnet (see Fig. 1f). We conjecture that the low-$T$ $C_m$ shows gapped behavior up to $T_c \sim 0.27$ K, $C_m \sim \exp(-\Delta_0/T)$ (see Fig. 1f), indicating gapped nature of the stripe state. The resulting gap, $\Delta_0 = 0.556(7)$ K, is of the same scale as the median value of $J_2^{zz}$.

All of the above underpins our interpretation of TmMgGaO$_4$ as the random $J_1^{zz}$ - $J_2^{zz}$ Ising antiferromagnet on the triangular lattice. Several effects going beyond this model are also worth mentioning. The discrepancies in the experimental and calculated magnetization process (Fig. 3a) likely indicate that randomness effects are not fully captured by the Lorentzian distribution of the microscopic parameters. Moreover, the susceptibility is measured to be finite, instead of zero, down to 30 mK, which may be due to regions where $J_1^{zz}$ and $J_2^{zz}$ significantly deviate from their median values. Although randomness can be deemed an excess intricacy of this system, it plays central role in the magnetism of TmMgGaO$_4$ by merging two singlet CEF levels of non-Kramers Tm$^{3+}$ into a quasidoublet and preserving Ising magnetism down to lowest temperatures.

Another interesting aspect is the possible deviation from purely Ising interactions. In non-Kramers ions, such as Pr$^{3+}$ in Pr$_2$TM$_2$O$_7$ (TM = Zr, Sn, Hf, and Ir), the weights of smaller angular moment states in the GS CEF (quasi) doublet contribute to superexchange interactions via quadrupole moments, generate non-Ising terms, and induce quantum fluctuations[19,20]. In TmMgGaO$_4$, the average effective pseudospin-1/2 $g$-factor of $g_\parallel = 13.18(1)$ (see Fig. 1a) is slightly lower than the upper limit of $2Jg_J = 14$. Thus, it is possible that the weights of smaller angular moment states in the GS CEF quasidoublet influence the low-temperature magnetism and trigger quantum fluctuations. Further insight into these effects can be obtained by studying magnetic excitations and dynamics via inelastic neutron scattering and muon spin relaxation, respectively.

At 0 T, a more generic pseudospin-1/2 Hamiltonian that is invariant under the $R\bar{3}m$ space group of TmMgGaO$_4$ is given by

$$\mathcal{H} = \sum_{\langle ij \rangle} [J_1^{zz} S_i^z S_j^z + J_1^{\pm}(S_i^+ S_j^- + S_i^- S_j^+) + J_1^{\pm\pm}(\gamma_{ij} S_i^+ S_j^+ + \gamma_{ij}^* S_i^- S_j^-)]$$

$$+ J_2^{zz} \sum_{\langle\langle kl \rangle\rangle} S_k^z S_l^z, \qquad (3)$$

where $S_i^{\pm} = S_i^x \pm i S_i^y$ is the time-reversal invariant quadrupole moment in the non-Kramers case[18-20,26,27], and $\gamma_{ij} = 1, \exp(i2\pi/3), \exp(-i2\pi/3)$ is the phase factor for the bond $ij$ along the **a**$_1$, **a**$_2$, **a**$_3$ direction, respectively (see Fig. 2a).

In conclusion, TmMgGaO$_4$ is an Ising antiferromagnet featuring the perfect triangular lattice of non-Kramer Tm$^{3+}$ ions that host robust Ising spins through the formation of the low-lying CEF quasidoublet as a result of structural randomness. Our comprehensive milli-Kelvin study reveals a weak NNN interaction, $J_2^{zz} \sim 0.09 J_1^{zz}$, which is large enough to release all the zero-point entropy expected in the NN TIAF model. We propose that below 0.27 K a frozen stripe state is formed in zero field, whereas field-induced states include the 1/3-plateau, 1/2-plateau, and fully spin-polarized phases. Further experiments on TmMgGaO$_4$ are feasible thanks to the availability of sizable single crystals and should address pending questions regarding the role of quantum fluctuations and the nature of spin excitations in the stripe and plateau phases of the triangular Ising antiferromagnet realized experimentally for the first time.

**Methods**

**Sample preparation.** Large and transparent single crystals (~ 1 cm) of TmMgGaO$_4$, Tm$_{0.04}$Lu$_{0.96}$MgGaO$_4$, and Yb$_{0.04}$Lu$_{0.96}$MgGaO$_4$ (Supplementary Figs 4, 6, and 8) were grown in a high-temperature optical floating zone furnace (FZ-T-10000-H-VI-VPM-PC, Crystal Systems Corp.)[9,17,18], using 53.0%, 60.7%, and 60.9% of the full power of the four lamps (the full power is 1.5 kW for each lamp), respectively. The single crystals were oriented by the Laue x-ray diffraction, and were cut consequently by a line cutter along the crystallographic *ab*-plane. The cut planes were cross-checked by both Laue (Supplementary Figs 4, 6, and 8) and conventional x-ray diffractions (Supplementary Figs 5, 7, and 9). The high-quality of the crystal was confirmed by the narrow reflection peaks, $2\Delta\Theta = 0.047 - 0.065°$ (FWHM).

**Sample characterization above 1.8 K.** The direct current (DC) magnetization ($1.8 \leq T \leq 400$ K and $0 \leq \mu_0 H \leq 7$ T) was measured by a magnetic property measurement system (MPMS, Quantum Design) using single crystals of $\sim 100$ mg. The DC magnetization up to 14 T was measured by a vibrating sample magnetometer in a physical property measurement system (PPMS, Quantum Design). The heat capacity ($1.8 \leq T \leq 400$ K and $0 \leq \mu_0 H \leq 12$ T) was measured using single crystals of $\sim 10$ mg in a PPMS. N-grease was used to facilitate thermal contact between the sample and the puck below 210 K, while H-grease was used above 200 K. The sample coupling was better than 99%. The contributions of the grease and puck under different external fields were measured independently and subtracted from the data.

**Millikelvin measurement below 2 K.** The heat capacity of the TmMgGaO$_4$, Tm$_{0.04}$Lu$_{0.96}$MgGaO$_4$, and Yb$_{0.04}$Lu$_{0.96}$MgGaO$_4$ single crystals was measured by a home-built setup in a $^3$He-$^4$He dilution refrigerator between 0.1 and 2.0 K at magnetic fields up to 5 T applied along the *c*-axis. Below $\sim 0.3$ K and in applied fields, the nuclear contribution becomes prominent, and the measured thermal relaxation slightly deviates from the two-tau model at short times[18]. We chose to exclude the 0.2 and 0.5 T heat capacity data below 0.12 and 0.2 K respectively, as the deviation is relatively large (adj. $R^2 < 0.9995$, see Ref. 18 for details). We fitted the 0, 0.2, and 0.5 T magnetic heat capacities using the function, $C_n(^{169}\Delta/T) + A\exp(-\Delta/T)$, from the lowest temperature up to the temperature of the minimum in $C_m$ (see Fig. 4a). Here $C_n(^{169}\Delta/T)$ is the nuclear heat capacity expressed by a two-level model, $^{169}\Delta$ and $\Delta$ are the nuclear and electronic spin gaps, respectively, and $A$ is a pre-factor[18]. The DC magnetization of TmMgGaO$_4$ between 0.024 and 2.0 K at magnetic fields up to 8 T applied along the *c*-axis, was measured by a high-resolution capacitive Faraday force magnetometer in a $^3$He-$^4$He dilution refrigerator[28]. The magnetic Grüneisen ratio or magnetocaloric effect, $\Gamma_m$ = $(dT/dH)/(\mu_0 T)$ = $-(dM_\parallel/dT)/C_p$, was measured by the alternating field technique ($\nu$ = 0.02 and 0.04 Hz) in a $^3$He-$^4$He dilution refrigerator[29,30].

**Exact calculation for the $J_1^{zz}$ - $J_2^{zz}$ TIAF model at 0 K.** The exact GS calculation for the TIAF model with only NN interaction had been performed up to a 56-site (8×7) cluster with periodic boundary conditions at 0 T. The zero-point entropy remains almost the same for larger-size calculations, $S_0^t$ = 44% − 50% $R\ln 2$ ($S_0^t$ = 47.3%, 46.8%, 50.0%, 47.6%, 43.5%, 47.0%, 46.5%, and 49.0% $R\ln 2$ for the 5×5, 6×5, 6×6, 7×6, 7×7, 8×7, 9×5, and 9×6 clusters with periodic boundary conditions, respectively), which is consistent with the previously reported results[5-7]. The exact GS calculation for the TIAF model with both NN and NNN interactions ($J_1^{zz} > J_2^{zz}$) had been performed on a 36-site (6×6) cluster with periodic boundary conditions. For the total $M_\parallel/M_\parallel^s$ = 0, 1/18, 2/18…1, the lowest system energies, $E(M_\parallel/M_\parallel^s)$, have been calculated respectively. The magnetization, $M_\parallel/M_\parallel^s$ in the longitudinal field (see Fig. 3a) was obtained by minimizing the function, $E(M_\parallel/M_\parallel^s)$-18$\mu_0 H_\parallel g_\parallel \mu_B(M_\parallel/M_\parallel^s)$. The calculated M-H curve shows no size-effects and no small steps (see Fig. 3a). Four different phases with stripe, 1/3-plateau, 1/2-plateau, and ferromagnetic spin correlations (see Fig. 2) are separated by three critical/transition fields, $\mu_0 H_{c1}$ = $6J_2^{zz}/(\mu_B g_\parallel)$, $\mu_0 H_{c2}$ = $3(J_1^{zz}-3J_2^{zz})/(\mu_B g_\parallel)$, and $\mu_0 H_{c3}$ = $3(J_1^{zz}+J_2^{zz})/(\mu_B g_\parallel)$, respectively. Our results are fully consistent with previous reports on the $J_1^{zz} > J_2^{zz}$ case[21,22].

**Data availability.** The data sets generated during and/or analysed during the current study are available from the corresponding author on request.

**Acknowledgements**

We thank Sebastian Esser for his technical help in Göttingen, and Yuanpai Zhou for his technical help in the calculation. The work was supported by the German Science Foundation through TRR-80 and the German Federal Ministry for Education and Research through the Sofja Kovalevskaya Award of the Alexander von Humboldt Foundation.

**Author contributions**

Y. L. synthesized and characterized the samples above 1.8 K. S. B. and Y. T. performed milli-K measurements. Y. L. analyzed the data and performed the calculations. Y. L. and A.A.T. wrote the manuscript with comments from all co-authors, P.G. supervised the project. The manuscript reflects the contributions of all authors.

**Competing interests**

The authors declare that they have no competing financial interests.

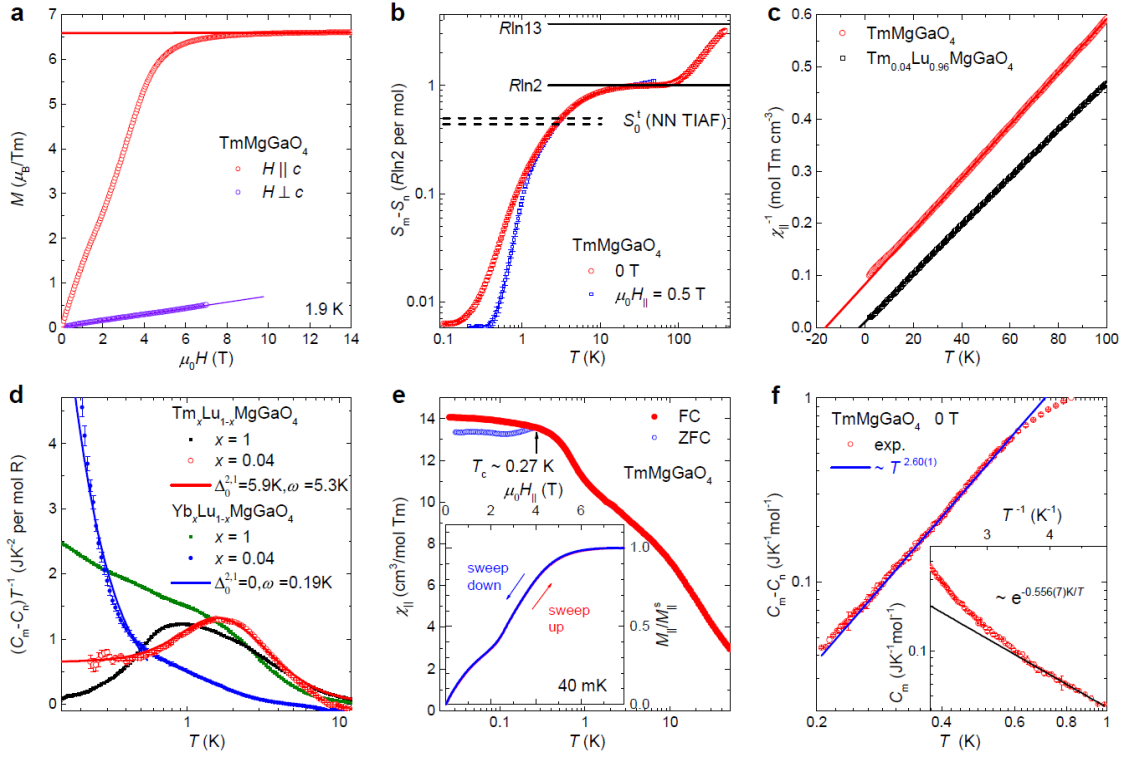

**Figure 1. Thermodynamic properties of TmMgGaO$_4$, Tm$_{0.04}$Lu$_{0.96}$MgGaO$_4$, YbMgGaO$_4$, and Yb$_{0.04}$Lu$_{0.96}$MgGaO$_4$ single crystals. a.** Magnetization of TmMgGaO$_4$ measured at 1.9 K in the fields parallel ($M_\parallel$) and perpendicular ($M_\perp$) to the $c$-axis. The red and violet lines show the linear fits to $M_\parallel$ above 10 T and to $M_\perp$, respectively. **b.** Magnetic entropy of TmMgGaO$_4$. **c.** Curie-Weiss fits to the susceptibilities of TmMgGaO$_4$ and Tm$_{0.04}$Lu$_{0.96}$MgGaO$_4$ along the $c$-axis. The data are corrected by the small constant Van Vleck susceptibility, $\chi_\parallel^{vv}$ = 0.003 cm$^3$/mol Tm, extracted from **a**. **d.** Magnetic heat capacities ($C_m$) of TmMgGaO$_4$, Tm$_{0.04}$Lu$_{0.96}$MgGaO$_4$, YbMgGaO$_4$, and Yb$_{0.04}$Lu$_{0.96}$MgGaO$_4$ at 0 T. The red and blue lines show, respectively, the fits to the data for Tm$_{0.04}$Lu$_{0.96}$MgGaO$_4$ and Yb$_{0.04}$Lu$_{0.96}$MgGaO$_4$ with Lorentzian distributions of $E_2$-$E_1$. **e.** Susceptibilities of TmMgGaO$_4$ measured under zero-field cooling and field cooling at 0.1 T with the field along the $c$-axis. The inset shows the magnetization measured at 40 mK. **f.** $C_m$ of TmMgGaO$_4$ with the blue line showing the power-law fit. The inset shows the corresponding $C_m$ vs. $T^{-1}$ plot with the black line showing the exponential (spin-gap) fit. The nuclear contributions ($\Delta S_n$(0.1 K) ~ 5.6%$R$ln2) have been subtracted (see main text).

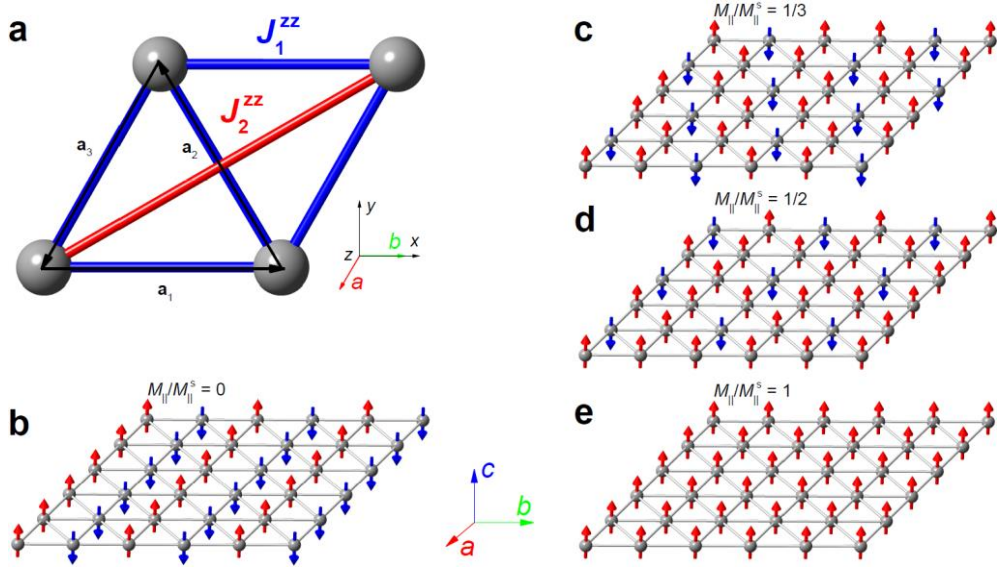

**Figure 2. Calculated result for the $J_1^{zz}$ - $J_2^{zz}$ TIAF model at 0 K. a.** TIAF model with both NN and NNN interactions. The *xyz*-coordinate system for the spin components is defined in the inset. **b.** Stripe phase observed at $H_\parallel < H_{c1}$, with $M_\parallel/M_\parallel^s = 0$. **c.** 1/3-plateau phase at $H_{c1} \leq H_\parallel < H_{c2}$ with $M_\parallel/M_\parallel^s = 1/3$. **d.** 1/2-plateau phase at $H_{c2} \leq H_\parallel < H_{c3}$ with $M_\parallel/M_\parallel^s = 1/2$. **e.** Fully spin-polarized phase at $H_\parallel \geq H_{c3}$ with $M_\parallel/M_\parallel^s = 1$.

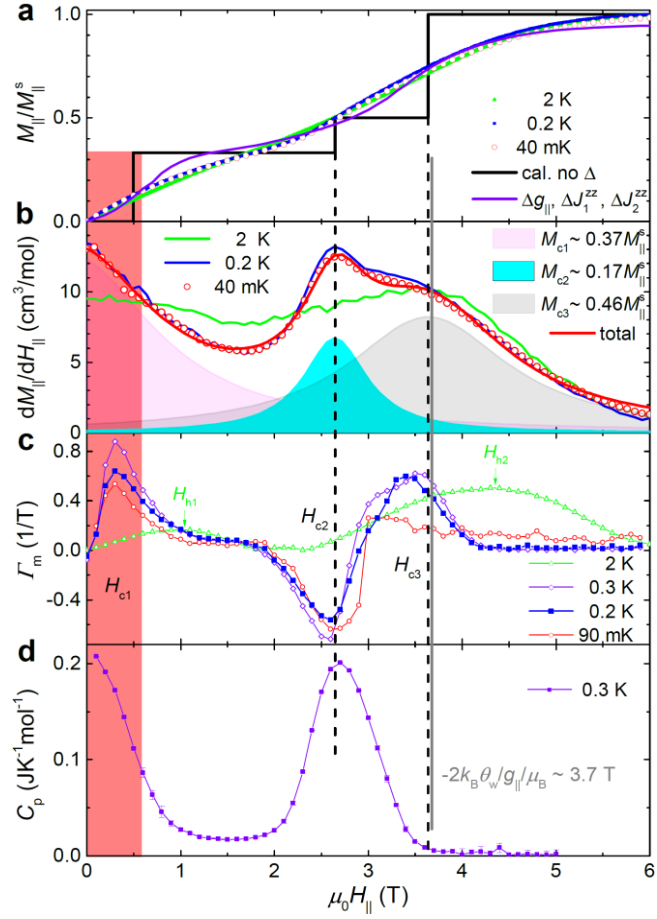

**Figure 3. Field dependence of the thermodynamic properties of TmMgGaO$_4$ in the field applied along the *c*-axis. a.** Magnetization ($M_{\|}/M_{\|}^s$) measured at 0.04, 0.2, and 2 K. The black line shows the calculation at 0 K without any randomness, and the violet line represents the least-square fit to the 0.04 K data with the Lorentzian distributions of $g_{\|}$, $J_1^{zz}$, and $J_2^{zz}$. **b.** Field dependence of the susceptibility (d$M_{\|}$/d$H_{\|}$) with the red line showing the three-peak Lorentzian fit. **c.** Field dependence of the magnetic Grüneisen ratio measured at 0.09, 0.2, 0.3, and 2 K. **d.** Field dependence of the heat capacity measured at 0.3 K.

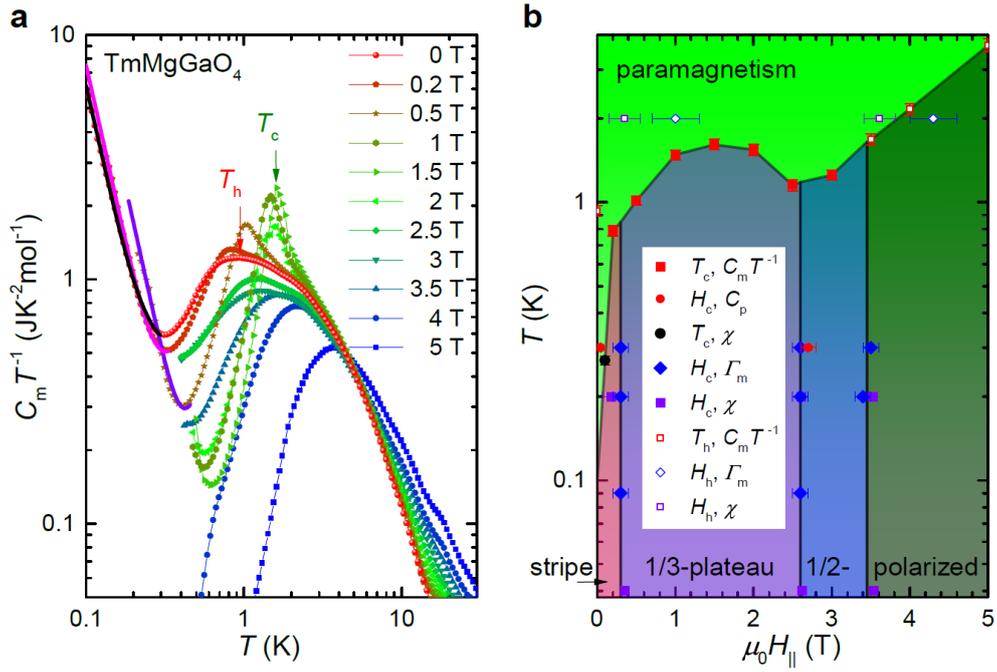

**Figure 4. Magnetic heat capacity and phase diagram of TmMgGaO$_4$. a.** Magnetic heat capacity measured at selected applied fields along the *c*-axis, without subtracting the nuclear contribution. The black, magenta, and violet lines show the nuclear & electronic spin heat-capacity fits at 0, 0.2, and 0.5 T, respectively (see main text). **b.** Phase diagram extracted from the heat capacity, magnetization, susceptibility, and magnetocaloric effect measurements (see **a**, Fig. 1, and Fig. 3).

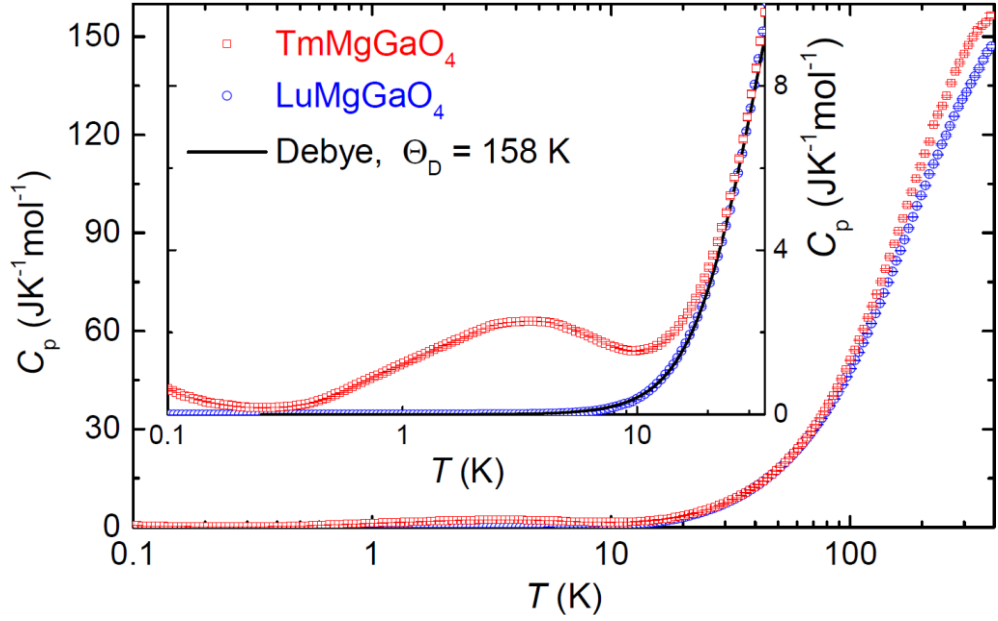

**Supplementary Figure 1. Heat capacity of the TmMgGaO$_4$ and LuMgGaO$_4$ single crystals measured at 0 T.** The inset presents a zoom-in plot of the low-$T$ data with the black line showing the Debye heat-capacity fit ($\theta_D = 158$ K).

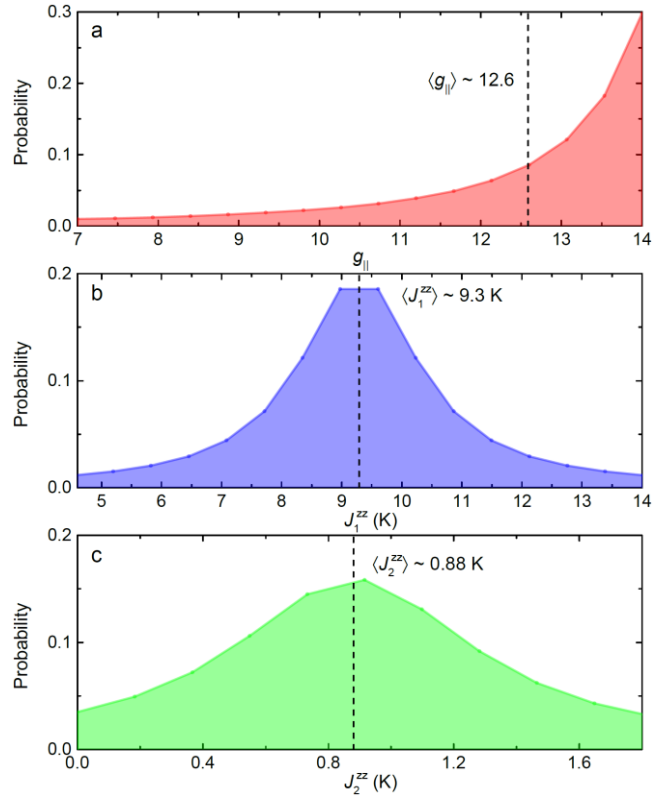

**Supplementary Figure 2. Fitted Lorentzian distributions of $g_\parallel$, $J_1^{zz}$, and $J_2^{zz}$ for TmMgGaO$_4$ (see main text). a.** Distribution of the effective g-factor, $g_\parallel$ ($g_\parallel \leq 2Jg_J = 14$). $\langle g_\parallel \rangle = 12.6$ and $\Delta g_\parallel = 1.5$ (full width at half maximum, FWHM) are obtained. **b.** Distribution of the nearest-neighbor interaction, $J_1^{zz}$ ($J_1^{zz} \geq 0$). $\langle J_1^{zz} \rangle = 9.3$ K and $\Delta J_1^{zz} = 2.4$ K (FWHM) are obtained. **c.** Distribution of the next-nearest-neighbor interaction, $J_2^{zz}$ ($J_2^{zz} \geq 0$). $\langle J_2^{zz} \rangle = 0.88$ K and $\Delta J_2^{zz} = 0.9$ K (FWHM) are obtained.

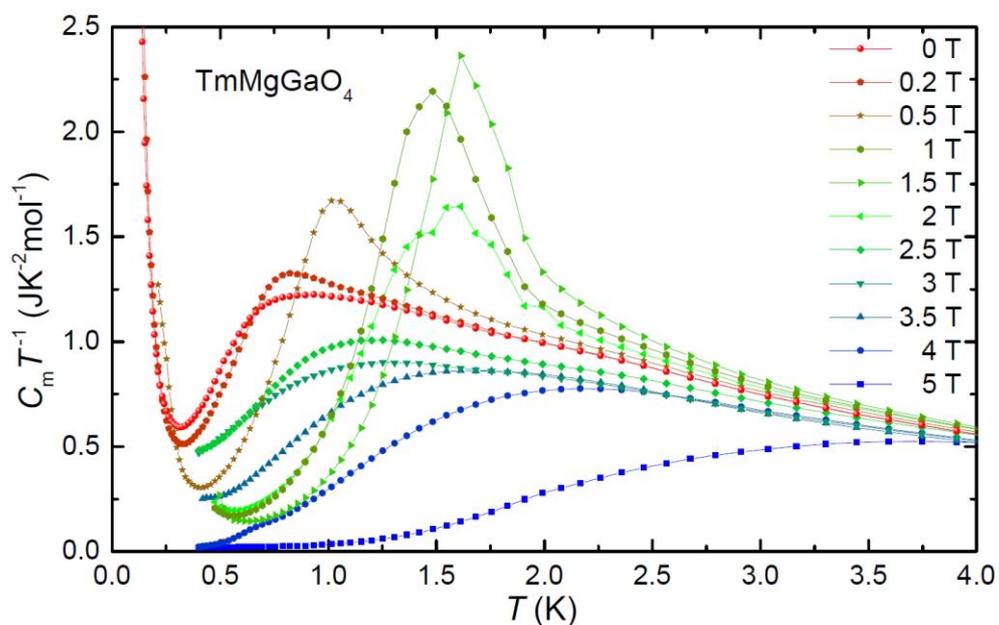

**Supplementary Figure 3. Magnetic heat capacity of TmMgGaO$_4$ measured at selected fields.** The phonon or lattice contribution was subtracted by the heat capacity of the non-magnetic LuMgGaO$_4$.

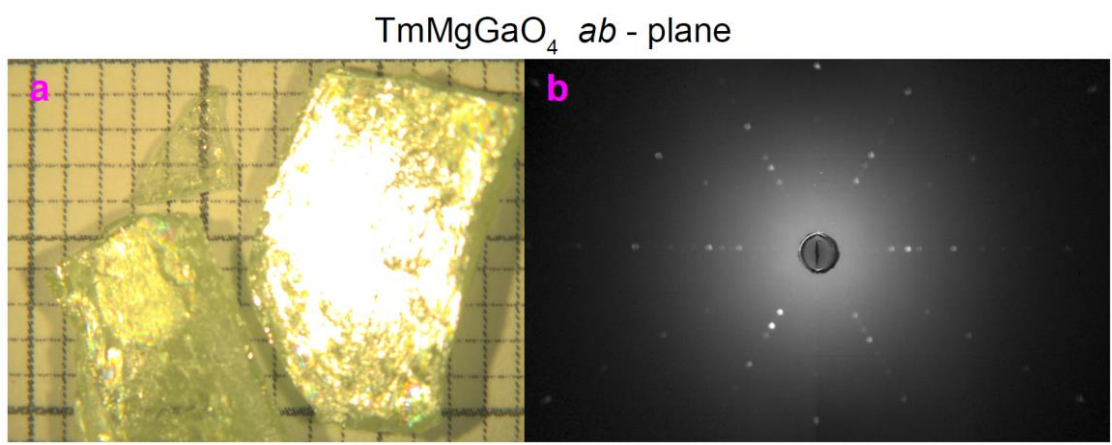

**Supplementary Figure 4. TmMgGaO$_4$ single-crystal sample. a.** Single crystals of TmMgGaO$_4$ cut along the *ab*-plane. **b.** Laue x-ray diffraction pattern on the *ab*-plane.

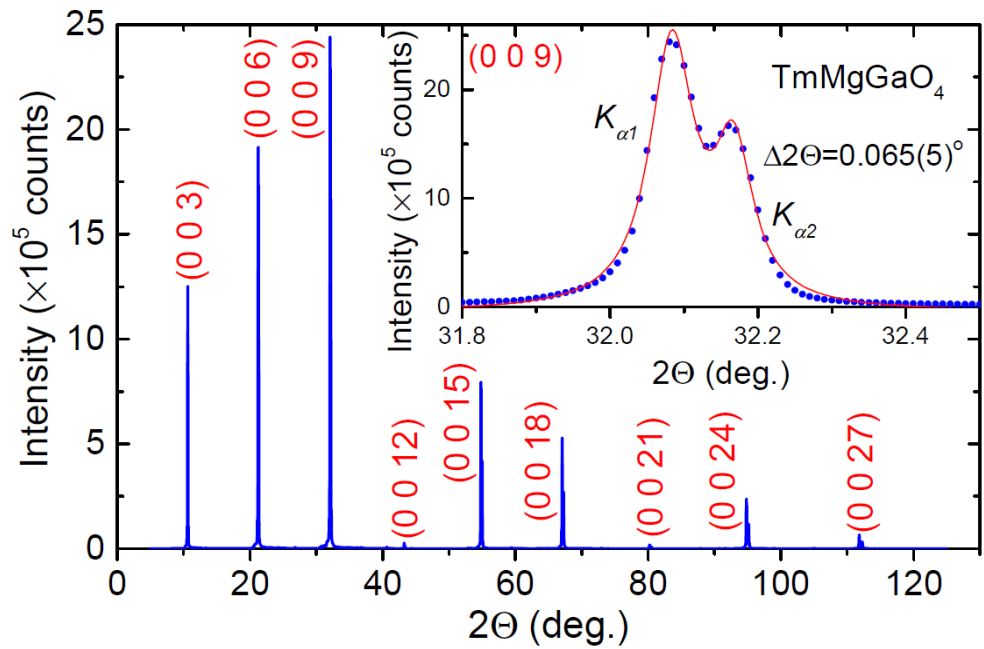

**Supplementary Figure 5. X-ray diffraction for the TmMgGaO$_4$ single crystal on the *ab*-plane.** The inset presents a zoom-in plot of the strongest Bragg peak, (0 0 9), where the angle (2θ) difference between the nearest-neighbor data points is 0.01°.

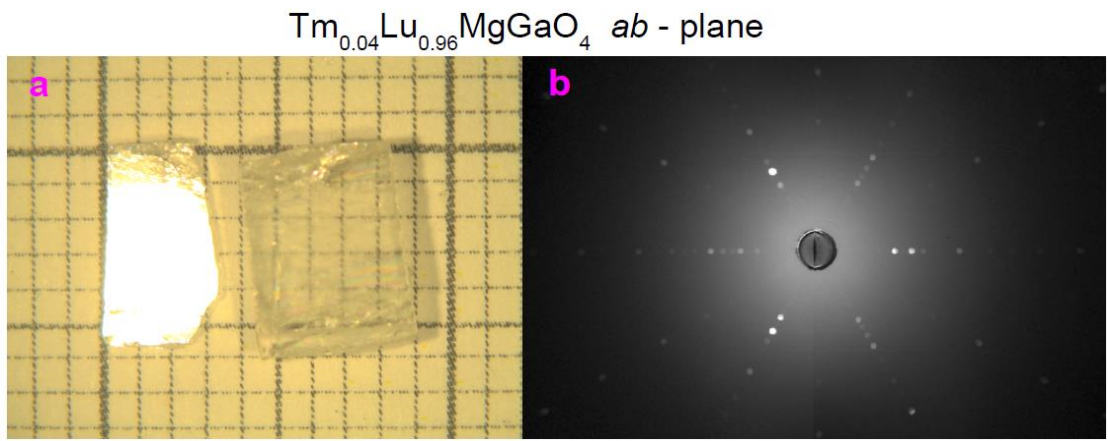

**Supplementary Figure 6. Tm$_{0.04}$Lu$_{0.96}$MgGaO$_4$ single-crystal sample. a.** Single crystals of Tm$_{0.04}$Lu$_{0.96}$MgGaO$_4$ cut along the *ab*-plane. **b.** Laue x-ray diffraction pattern on the *ab*-plane.

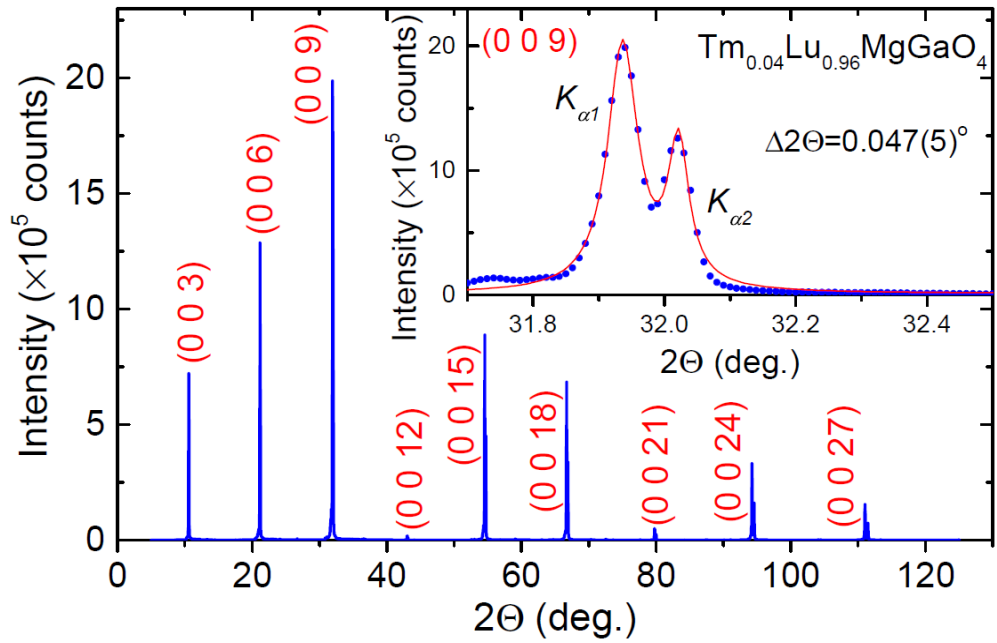

**Supplementary Figure 7. X-ray diffraction for the $Tm_{0.04}Lu_{0.96}MgGaO_4$ single crystal on the ab-plane.** The inset presents a zoom-in plot of the strongest Bragg peak, (0 0 9), where the angle (2θ) difference between the nearest-neighbor data points is 0.01°.

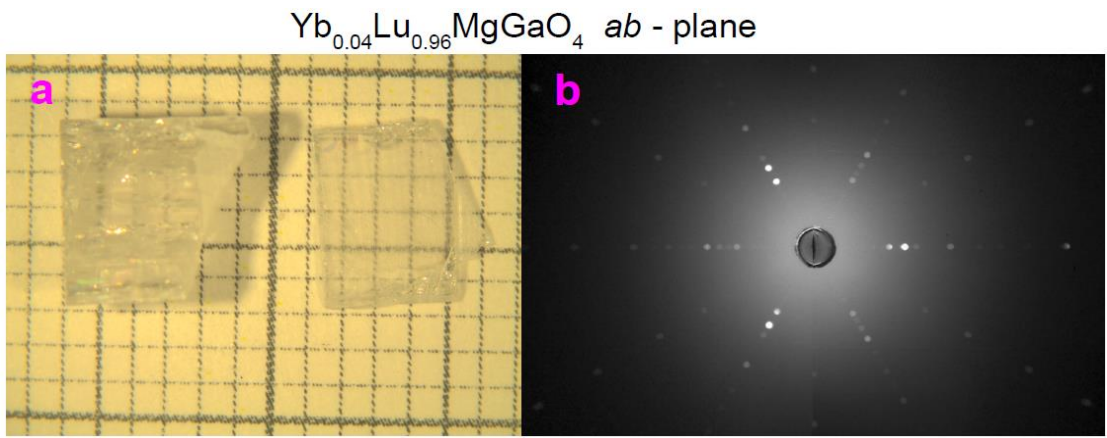

**Supplementary Figure 8. $Yb_{0.04}Lu_{0.96}MgGaO_4$ single-crystal sample. a.** Single crystals of $Yb_{0.04}Lu_{0.96}MgGaO_4$ cut along the *ab*-plane. **b.** Laue x-ray diffraction pattern on the *ab*-plane.

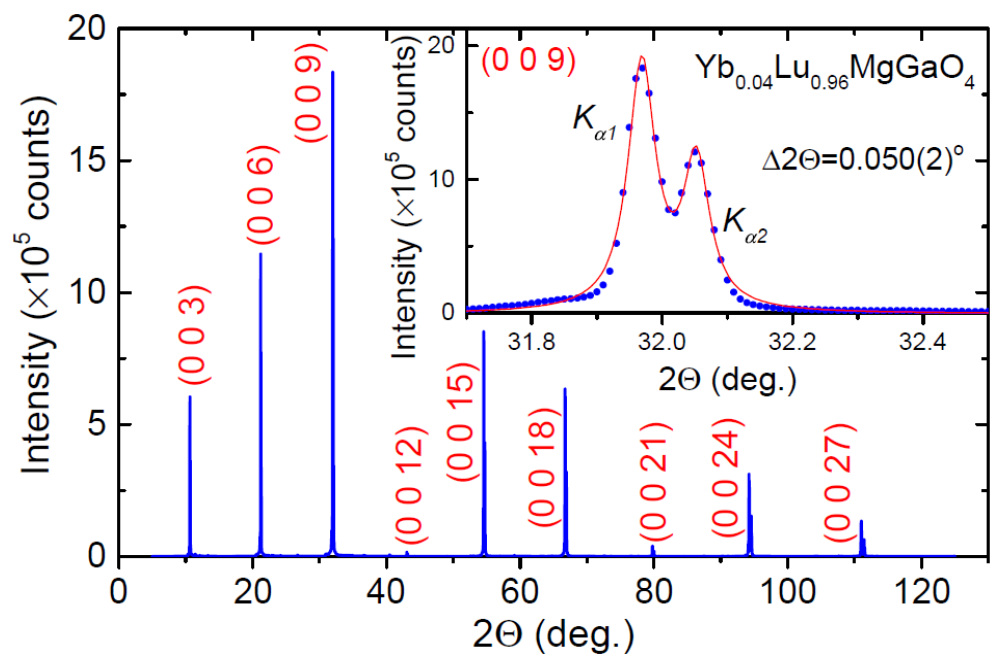

**Supplementary Figure 9. X-ray diffraction for the $Yb_{0.04}Lu_{0.96}MgGaO_4$ single crystal on the *ab*-plane.** The inset presents a zoom-in plot of the strongest Bragg peak, (0 0 9), where the angle (2θ) difference between the nearest-neighbor data points is 0.01°.